\documentclass[a4paper]{article}

\usepackage{INTERSPEECH2021}
\usepackage{graphicx}
\usepackage{float}
\usepackage{booktabs}
\usepackage{xcolor}
\usepackage{color}
\usepackage{hyperref}

\usepackage{subcaption}

\title{{DelightfulTTS: The Microsoft Speech Synthesis System for \\ Blizzard Challenge 2021}}
\name{Yanqing Liu$^1$, Zhihang Xu$^1$, Gang Wang$^1$, Kuan Chen$^1$, Bohan Li$^1$, Xu Tan$^2$, Jinzhu Li$^1$, Lei He$^1$, Sheng Zhao$^1$}
\address{
  $^1$Microsoft Azure Speech, China\\
  $^2$Microsoft Research Asia, China}
\email{\{yanqliu, zhihangxu, gaw, kuan.chen, bohli, xuta, jinzl, helei, szhao\}@microsoft.com}

\begin{document}

\maketitle

\begin{abstract}
This paper describes the Microsoft end-to-end neural text to speech (TTS) system: DelightfulTTS for Blizzard Challenge 2021. The goal of this challenge is to synthesize natural and high-quality speech from text, and we approach this goal in two perspectives: The first is to directly model and generate waveform in 48 kHz sampling rate, which brings higher perception quality than previous systems with 16 kHz or 24 kHz sampling rate; The second is to model the variation information in speech through a systematic design, which improves the prosody and naturalness. Specifically, for 48 kHz modeling, we predict 16 kHz mel-spectrogram in acoustic model, and propose a vocoder called HiFiNet to directly generate 48 kHz waveform from predicted 16 kHz mel-spectrogram, which can better trade off training efficiency, modelling stability and voice quality. We model variation information systematically from both explicit (speaker ID, language ID, pitch and duration) and implicit (utterance-level and phoneme-level prosody) perspectives~\cite{tan2021survey}: 1) For speaker and language ID, we use lookup embedding in training and inference; 2) For pitch and duration, we extract the values from paired text-speech data in training and use two predictors to predict the values in inference; 3) For utterance-level and phoneme-level prosody, we use two reference encoders to extract the values in training, and use two separate predictors to predict the values in inference. Additionally, we introduce an improved Conformer~\cite{gulati2020conformer} block to better model the local and global dependency in acoustic model~\cite{ren2020fastspeech, chen2021adaspeech}. For task SH1, DelightfulTTS achieves 4.17 mean score in MOS test and 4.35 in SMOS test, which indicates the effectiveness of our proposed system.

\end{abstract}

\section{Introduction}
The Blizzard Challenge\footnote{\url{https://www.synsig.org/index.php/Blizzard_Challenge}} is aimed at advancing the technologies in text to speech (TTS)~\cite{shen2018natural, ren2019fastspeech, ren2020fastspeech} by comparing and understanding different approaches, and has been organized annually since 2005 \cite{black2005blizzard}. The basic task is to build high-quality TTS systems based on the speech database provided by the organizers. Participants use their developed systems to synthesize audio from the given test set; the generated audio samples are used to evaluate performance of different system through subjective listening tests and objective metrics.

Typical TTS systems consist of three key components~\cite{tan2021survey}: text analysis, acoustic model, and vocoder. Firstly, input text is normalized and transformed into linguistic features of different levels (e.g., phoneme level, syllable level, word level) through a text analysis (TTS front-end) module. Then, these linguistic features are transformed into an intermediate acoustic representation like mel-spectrogram by an acoustic model. At last, the acoustic representation is converted to waveform with a vocoder. Rich unit selection based concatenative speech synthesis \cite{hunt1996unit} and statistical parametric speech synthesis like HMM/DNN/LSTM based models \cite{tokuda2000speech, ze2013statistical, fan2014tts}, have been the most popular methods in the past. Recently, end-to-end neural TTS has been researched extensively. Tacotron 2 \cite{shen2018natural} introduced an attention-based encoder-decoder acoustic model to predict mel-spectrogram given a character sequence and a WaveNet model \cite{oord2016wavenet} to synthesize waveform from mel-spectrogram, which achieves high voice quality but suffers from robustness issues and slow speed in both training and inference. To improve training speed, TransformerTTS \cite{li2019neural} has been proposed to synthesize high-quality speech with Transformer \cite{vaswani2017attention}. \cite{li2020robutrans} tried to solve the stability problems like repeated phonemes by using linguistic features and phoneme duration with autoregressive Transformer, but the inference is still slow due to autoregressive nature. FastSpeech \cite{ren2019fastspeech} leads a research trend on non-autoregressive TTS, by using feed-forward transformer in encoder and decoder for parallel generation and a duration predictor for length duration, which is much faster and more stable than autoregressive models like Tacotron 2 \cite{shen2018natural}. 

While there is much progress in the research on TTS and the quality of synthesized speech has been impressive, the perception quality and naturalness still have large space to improve. 
With this goal in mind, we participate the Blizzard Challenge 2021, and design an end-to-end neural text to speech (TTS) system called DelightfulTTS to advance the perception quality and naturalness. We approach this goal from two aspects: 1) The bandwidth to convey high-quality speech should be enough. We model and generate 48 kHz waveform, since high sampling rate (48 kHz) can have larger range of frequency to convey expression and prosody~\cite{chen2020hifisinger} and thus can result in higher perception quality than 16 kHz or 24 kHz. 2) The speech itself should be expressive. We model variation information~\cite{tan2021survey} in speech through a systematic design to improve prosody and naturalness. TTS is a typical one-to-many mapping problem where there could be multiple varying speech outputs (e.g., different pitch, duration, speaker, prosody, emotion, etc) for a given text input. Modeling these variation information~\cite{stanton2018predicting, ren2020fastspeech, sun2020generating, chen2021adaspeech} can mitigate the one-to-many mapping problem and thus improve expressiveness and fidelity of synthesized speech. We describe the specific design in each aspect as follows:

\begin{itemize}
\item To generate 48 kHz waveform, typical method is to generate 48 kHz mel-spectrogram and then generate 48 kHz waveform from the generated 48 kHz mel-sepctrogram. However, according to the recent advances in TTS~\cite{tan2021survey, kong2020hifi}, the quality of the waveform generated by vocoder nearly matches to that of the recording~\cite{kong2020hifi}, while the quality of the mel-spectrogram generated by acoustic model still has gap to recording due to the difficulty of prosody modeling. Intuitively speaking, vocoder only need to recover phase information given ground-truth spectrogram, while acoustic model needs to close the large gap between text and spectrogram, such as duration, pitch, formant, voiced/unvoiced part, prosody, etc. When increasing the sampling rate to 48 kHz, the burden of acoustic model is further increased. According to the “cannikin law (wooden barrel principle)", the capacity of a wooden barrel is determined by its shortest stave. Thus, to relieve the burden on acoustic model, we predict 16 kHz mel-spectrogram in acoustic model, and generate 48 kHz waveform from predicted 16 kHz mel-spectrogram using our recently developed HiFiNet~\cite{ms2020hifinet} vocoder.

\item Nevertheless, the burden of acoustic model is still heavy due to one-to-many mapping. Thus, we model variation information systematically from both explicit (speaker ID, language ID, pitch, and duration) and implicit (utterance-level and phoneme-level prosody) perspectives~\cite{tan2021survey}: 1) For speaker and language ID, we use lookup embedding in training and inference; 2) For pitch and duration, we extract the values from paired text-speech data in training and use two predictors to predict the values in inference; 3) For utterance-level and phoneme-level prosody, we use two reference encoders to extract the values in training, and use two separate predictors to predict the values in inference. We build the acoustic model based on non-autoregressive generation model AdaSpeech~\cite{chen2021adaspeech} and FastSpeech~\cite{ren2019fastspeech}, with an improved Conformer module~\cite{gulati2020conformer} to better model the local and global dependency in mel-spectrogram.

\end{itemize}

According to the official evaluation of Blizzard Challenge 2021 in task SH1, our proposed system achieves 4.17 mean score in MOS test and 4.35 in SMOS test (recording has 4.21 and 4.07 mean score in MOS and SMOS test respectively), which is better than all other participants and matches to the quality of recording in terms of mean score perspective of MOS and SMOS, which demonstrate the superiority of our system to advance the perception quality and naturalness\footnote{Audio samples: \url{https://cognitivespeech.github.io/delightfultts}}.

The rest of the paper is organized as follows. Section 2 presents the architecture of the proposed system. Section 3 describes the task, data process and training strategy. Section 4 introduces the subjective evaluation results. Finally, conclusion and future work are presented in Section 5.

\section{Our Approach}

\subsection{System Overview}

As shown in Figure 1a,  our proposed system consists of an acoustic model that generates mel-spectrogram from text and a vocoder that generates waveform from mel-spectrogram. To ensure high fidelity and quality of the synthesized speech, we directly generate waveform in 48 kHz sampling rate, since it can have larger range of frequency to convey expression and prosody~\cite{chen2020hifisinger}. To better trade off the difficulty in acoustic model and vocoder, we predict 16 kHz mel-spectrogram with acoustic model, and generate 48 kHz waveform from predicted 16 kHz mel-spectrogram with vocoder, which are described in Section~\ref{sec_tradeoff}. The acoustic model consists of an improved Conformer module~\cite{gulati2020conformer} in the phoneme side and mel-spectrogram side respectively, and a variance adaptor in between the two conformer modules, as shown in Figure 1b. We model the variance information through variance adaptor to improve the prosody and naturalness in both explicit and implicit ways (Figure 1c), as introduced in Section~\ref{sec_var}. We introduce the details of the improved Conformer module (Figure 1d) in Section~\ref{sec_cfm}. The vocoder is based on our previously proposed HiFiNet~\cite{ms2020hifinet}. We introduce each design in our proposed system in the following subsections.

\subsection{Tradeoff between Acoustic Model and Vocoder}
\label{sec_tradeoff}
We directly generate waveform in 48 kHz sampling rate to ensure high fidelity and quality for speech synthesis.
There is a tradeoff in the task difficulty between acoustic model and vocoder, which influences the design choices on acoustic model and vocoder. Intuitively speaking, the burden of acoustic model is larger than that of vocoder, according to several perspectives: 1) Information gap. Vocoder only needs to recover phase information given spectrogram, while acoustic model needs to close the large gap between text and spectrogram, such as duration, pitch, formant, voiced/unvoiced part, prosody, etc. Thus, the large information gap between text and spectrogram makes the task of acoustic model more difficult than that of vocoder. 2) Current research progress. According to the recent advances in TTS~\cite{tan2021survey, kong2020hifi}, the quality of vocoding task nearly matches to that of the recording~\cite{kong2020hifi}, while the acoustic modeling task still has large improvement space due to the difficulty of prosody modeling. 3) Model capacity. We conduct a brief survey on the size of model parameters of the mainstream acoustic models and vocoders (mainly GAN-based)~\cite{tan2021survey}, and find that the parameter sizes of acoustic models are usually 5x (or more) larger than that of vocoders.

If we generate 48 kHz mel-spectrogram with acoustic model and then generate 48 kHz waveform with vocoder, the task difficulties of both acoustic model and vocoder are increased, compared with 16 kHz counterparts. However, the burden of acoustic model is even larger according to the analysis above. According to the “cannikin law” (i.e., “wooden barrel principle”), the capacity of a wooden barrel is determined by its shortest stave. In our case, the quality of speech synthesis would be constrained by the acoustic model if not well handled. Thus, to relieve the burden on acoustic model, we predict 16 kHz mel-spectrogram with acoustic model, and generate 48 kHz waveform from predicted 16 kHz mel-spectrogram using our recently developed HiFiNet \cite{ms2020hifinet} vocoder, which is optimized with a generator and discriminator-based network \cite{goodfellow2014generative}. In this way, the task difficulty of acoustic model is reduced and that of vocoder is slightly increased (with both phase reconstruction and super-resolution), resulting in a more balanced way.

\begin{figure*}[h]
  \centering
  \includegraphics[width=0.93\textwidth,trim=0.5cm 0.5cm 0.5cm 1.0cm,clip=true]{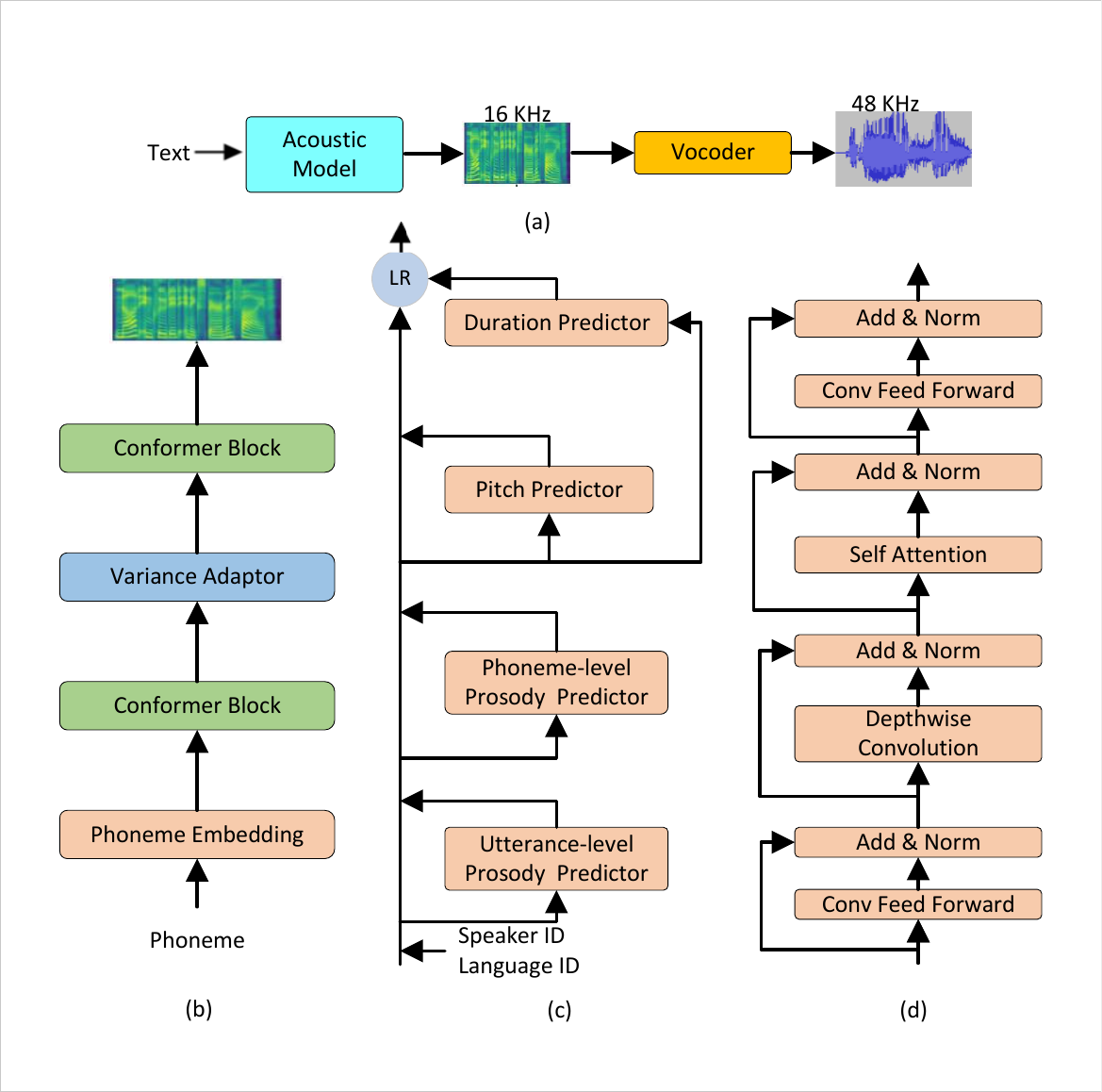}
  \caption{Overview of our proposed TTS system. Figure (a) shows the overall pipeline for DelightfulTTS. Figure (b) shows the architecture of the acoustic model. Figure (c) shows the variance adaptor with explicit and implicit variation information modeling. Figure (d) shows the improved Conformer block. }
  \label{fig:v16}
\end{figure*}

\subsection{Variation Information Modelling}
\label{sec_var}
TTS is a typical one-to-many mapping problem where there could be multiple varying speech outputs (e.g., different pitch, duration, speaker, prosody, emotion, etc) for a given text input. It is critical to model these variation information in speech so as to improve expressiveness and fidelity of synthesized speech. While previous works~\cite{stanton2018predicting, ren2020fastspeech, sun2020generating} have tried different methods to model the information, they focus on a specific aspect and cannot model in a comprehensive and systematic way. In this paper, considering that different variation information can be complementary to each other, we propose a unified way to model them in the proposed variance adaptor (as shown in Figure 1c). Observing that some variation information can be obtained implicitly (e.g., pitch can be extracted by some tools) or explicitly (e.g., utterance-level prosody can only be learned by the model), we categorize all the information we model as follows: 1) Explicit modeling, includes language ID, speaker ID, pitch, duration. 2) Implicit modeling, including utterance-level and phoneme-level prosody.

For speaker and language ID, we use lookup embedding in training and inference. For pitch and duration, we extract the values from paired text-speech data in training and use two predictors~\cite{ren2020fastspeech} to predict the values in inference. For utterance-level and phoneme-level prosody, we use two reference encoders to extract the values in training~\cite{chen2021adaspeech} and use two separate predictors to predict the values in inference. The two reference encoders are both made up of convolution and RNN layers. Utterance-level prosody vector is obtained by the last RNN hidden and a style token layer~\cite{wang2018style}. Phoneme-level prosody vectors are obtained by using the outputs of phoneme encoder (phoneme-level) as query to attend to the outputs of mel-spectrogram reference encoder (frame-level). Different from \cite{elias2021parallel}, we do not use VAE, but directly use the latent representation as phoneme-level vector for training stability~\cite{chen2021adaspeech}. The utterance-level prosody predictor contains a GRU layer followed by a bottleneck module to predict the prosody vector. The phoneme-level prosody predictor takes both the outputs of text encoder and utterance-level prosody vector as input. With the help of utterance-level prosody vector, we do not need an autoregressive prosody predictor as \cite{elias2021parallel} for faster inference. By unifying explicit and implicit information in different granularities (language-level, speaker-level, utterance-level, phoneme-level)~\cite{tan2021survey} in the variance adaptor, we can achieve better expressiveness in prosody and controllability in pitch and duration.

\subsection{Conformer in Acoustic Model}
\label{sec_cfm}
Conformer \cite{gulati2020conformer} is a Transformer variant that integrates both CNNs and Transformers components. Originally used for end-to-end speech recognition, it exhibits better accuracy with fewer parameters than previous work on several open ASR datasets, and achieves a new state-of-the-art performance. The multi-headed self-attention (MHSA) is conformer integrates an important technique from Transformer-XL\cite{dai2019transformer} with relative sinusoidal positional encoding scheme. Conformer proposes a novel combination of self-attention and convolution, in which self-attention learns the global interaction while the convolutions efficiently capture the local correlations. We think that the global and local interactions are especially important for TTS considering it has a longer output sequence than machine translation or speech recognition in decoder. For non-autoregressive TTS model, a powerful modelling unit is critical because each frame in decoder cannot see its history as autoregressive model does. In our proposed system, we make several improvements to the original Conformer architecture. First, we replace Swish~\cite{ramachandran2017searching} with ReLU and observe better generalization especially on long sentences. Second, we switch the order of self-attention and depthwise convolution for faster convergency. Third, we replace the linear layers in feed forward-modules of Macoron \cite{lu2019understanding} structure with convolution layers, which results in better prosody and audio fidelity. Altogether, our improved conformer block is composed of four modules stacked together, i.e., a convolutional feed-forward module, a depthwise convolution module, a self-attention module and a second convolutional feed-forward module in the end, illustrated in Figure 1d. 

\section{Task Description, Data Processing, and Training Strategy}

\subsection{Task Description}

The task in Blizzard Challenge 2021 we joined is Hub task 2021 - SH1. There're about 5 hours of speech data with 48 kHz sampling rate from a female native speaker of European Spanish. The hub task is to build a voice from the provided European Spanish data to synthesize texts containing only Spanish words.

\subsection{Data Processing}
Phoneme representation is widely used as the input of end-to-end TTS systems like recent neural TTS models \cite{li2019neural, li2020robutrans, ren2019fastspeech, li2020moboaligner}, which can avoid ambiguity depending on the text context. In this task we leverage an internal text preprocess module that converts the input Spanish text into a phoneme sequence. First, we split the text into sentences through sentence separation module. Second, we perform text normalization (TN) with a rule-based TN module. Third, we convert the normalized text to phoneme sequence with a G2P \cite{bali2004tools} model and a Spanish lexicon. 
The duration target is extracted by an internal force alignment model. For acoustic model training, audio used in our system is down-sampled to 16 kHz. Then mel-spectrograms are computed through a short time Fourier transform (STFT) using a 50 ms frame size, 12.5 ms frame hop, and a Hann window function. For vocoder training, the original 48k audios were used in the training of HiFiNet \cite{ms2020hifinet}.

\subsection{Training Strategy}
The overall loss function for the proposed acoustic model is
\begin{equation}
L = L_{utt}+L_{phone}+L_{pitch}+L_{dur}+L_{iter}+L_{ssim}
\end{equation}
where $L_{utt}$ is the L1 loss between predicted utterance-level prosody vector and the vector extracted from utterance-level reference encoder, $L_{phone}$ is the L1 loss between the predicted phoneme-level prosody vectors and the vectors extracted from phoneme-level reference encoder, $L_{pitch}$/$L_{dur}$ is the L1 loss between predicted pitch/duration and the ground-truth pitch/duration. For better convergence and voice quality, the output of each Conformer block in the mel-spectrogram side is projected into 80-dimension mel-spectrogram respectively, which is used to calculate mel-spectrogram loss with the ground-truth mel-spectrogram. Thus, $L_{iter}$ is the sum of mel-spectrogram L1 loss between the predicted and ground-truth mel-spectrogram in each Conformer block~\cite{elias2021parallel}. For better audio fidelity, we use SSIM (Structural SIMilarity)~\cite{wang2004image} to measure the similarity between predicted and ground-truth mel-spectrogram in the final Conformer block, denoted as $L_{ssim}$. We use the predicted mel-spectrogram in the final Conformer block as the final mel-spectrogram prediction. Besides the provided 5 hours data, we also use an extra internal dataset from Microsoft, which contains about 40 hours European Spanish data and 40 hours Mexican Spanish data. Acoustic model and vocoder are both pretrained on the full dataset and then finetuned on the provided single speaker data. The Conformer module in both phoneme and mel-spectrogram sides consist of 6 improved Conformer blocks, with attention dimension 384 and hidden size of convolutional feed-forward module 1536. The kernel size in the depthwise convolution module is 7. We train DelightfulTTS on 4 NVIDIA V100 GPUs and each GPU has a batch size of about 6000 speech frames.

\section{Results}

The evaluation of Blizzard Challenge 2021 includes four aspects: including naturalness test with MOS (mean opinion score), similarity test with SMOS (similarity mean opinion score), SUS intelligibility test with WER (word error rate) and Sharvard intelligibility test with WER. In MOS part, listeners listened to one sample and chose a score which represented how natural or unnatural the sentence sounded on a scale of 1 to 5. For SMOS they chose a score that represented how similar the synthetic voice sounded to the voice in the reference samples on a scale from 1 to 5. In intelligibility test, listeners heard one utterance in each part and typed in what they heard. Listeners were allowed to listen to each sentence only once. The synthesized and natural audios were carefully rated by three types of listeners who are involved by paid listeners, online volunteers, and speech experts. Our system is denoted as system F, and natural speech recorded by the original speaker is denoted as System R. Overall, the naturalness (MOS) of system F is significantly higher than all other systems, and is not significantly different to natural speech; speaker similarity (SMOS) of system F is better than all other systems.

\subsection{Naturalness Test}

Naturalness test evaluation results of all systems is showed in Figure 2, scored by all listeners. In synthesized systems, the naturalness (MOS) of system F (mean score 4.17) is significantly higher than all other systems, and is not significantly different to natural speech(mean score 4.21), while mean score of others is below 4.0. This shows the superiority of our DelightfulTTS over other systems.

\subsection{Similarity Test}

Figure 3 presents the mean opinion score of similarity evaluation results for all systems, scored by all listeners. Speaker similarity (SMOS) of system F is better than all other systems. Our system F is 4.35 while recording R is 4.07, other systems are all below 4.0.

\subsection{Intelligibility of Sentences on SUS}

Intelligibility test result of SUS set is presented in Figure 4 by paid listeners. The SUS \cite{benoit1996sus} set for this test was manually generated using the grammar structures. In this section, the result shows that our proposed system F was significantly better than system L and N, and no systems were significantly better than system F.

\subsection{Intelligibility of Sentences (INT) on Sharvard}

The sentences for this section came from the Sharvard corpus \cite{aubanel2014sharvard}. Note that this means the natural recordings are of a different speaker to that used to build the TTS systems. Presented in Figure 5, the results show that there were no significant differences between system F and other systems. 
\begin{figure}[h]
  \centering
  \includegraphics[width=\linewidth]{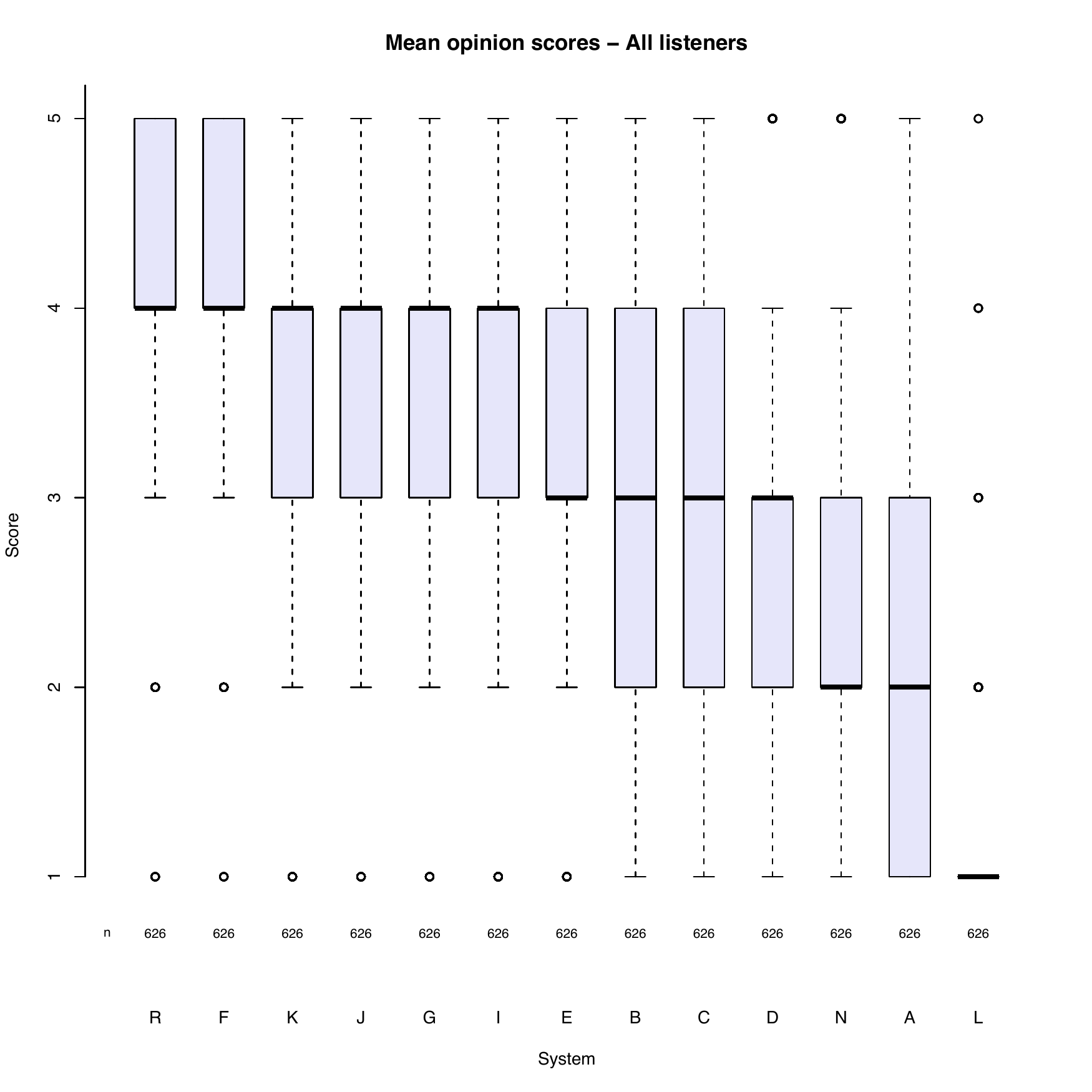}
  \caption{Boxplot of naturalness scores of each submitted system for all listeners. Our System is identified as F.}
  \label{fig:mos_all}
\end{figure}

\begin{figure}[h]
  \centering
  \includegraphics[width=\linewidth]{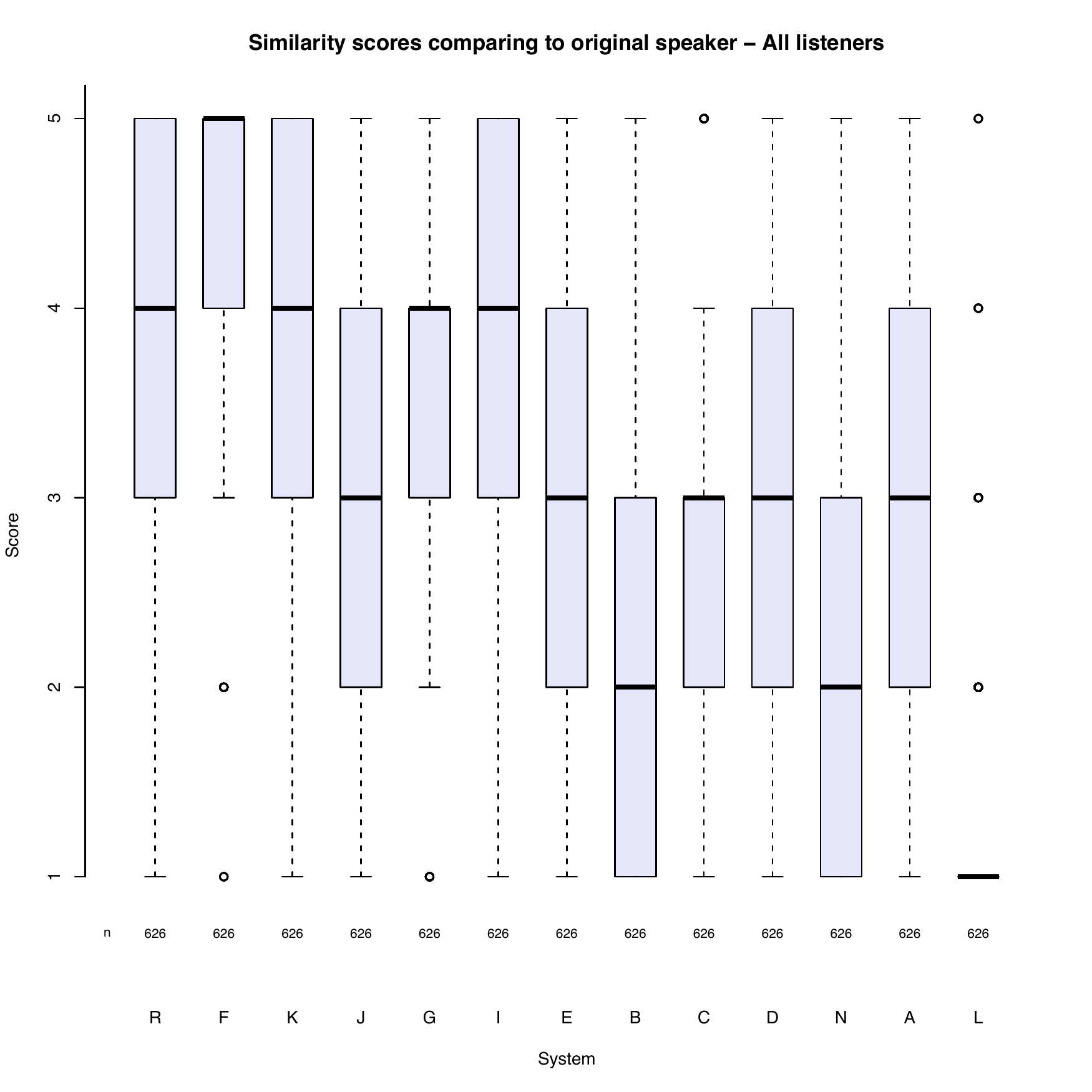}
  \caption{Similarity scores compared to original speaker. Our System is identified as F.}
  \label{fig:mos_all}
\end{figure}

\section{Conclusions}

This paper presents the Microsoft speech synthesis system for Blizzard Challenge 2021. We approach the goal of synthesizing natural and high-quality speech from text in two aspects: The first is to directly model and generate waveform in 48 kHz sampling rate, with a good tradeoff of the task difficulties between the acoustic model and the HiFiNet vocoder, which brings higher perception quality than previous systems with lower sampling rate; The second is to model the variation information in speech through a systematic design including both explicit and implicit modeling, which improves the prosody and naturalness. 

\begin{figure}[H]
  \centering
  \includegraphics[width=\linewidth]{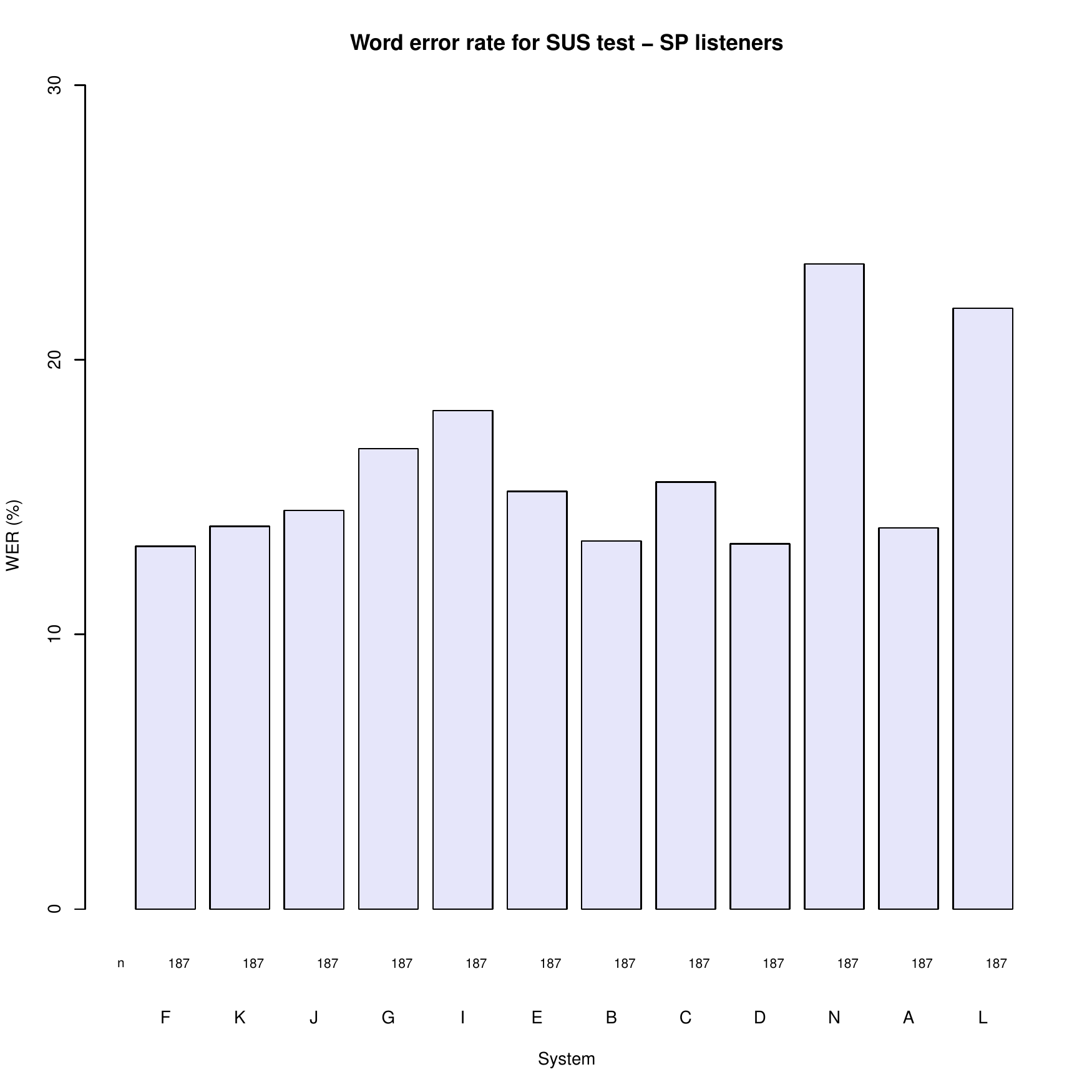}
  \caption{Word error rates for SUS test.}
  \label{fig:int_sus11}
\end{figure}

Overall, the naturalness (MOS) of system F is significantly higher than all other systems, and is not significantly different to natural speech; speaker similarity (SMOS) of system F is better than all other systems, which demonstrate the superiority of our proposed system. For future work, we will further improve the variation information modeling on multi-speaker multi-lingual multi-style datasets and investigate the style transfer capability across different languages and speakers.
\begin{figure}[H]
  \centering
  \includegraphics[width=\linewidth]{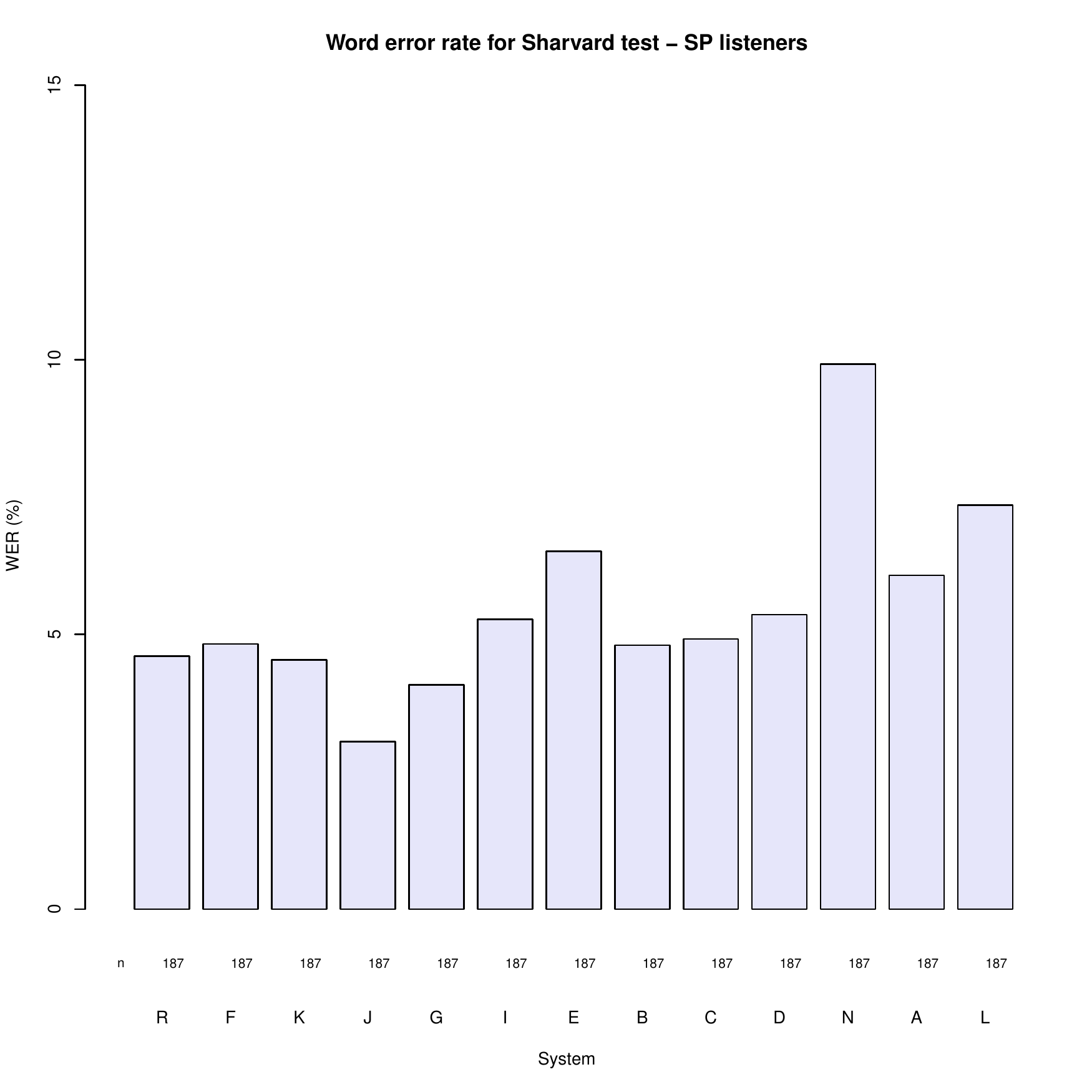}
  \caption{Word error rates for Sharvard test.}
  \label{fig:int_sha}
\end{figure}

\bibliographystyle{IEEEtran}

\bibliography{mybib}

\end{document}